\documentclass{cicc}
\firstpage{1}
 \articletype{Communication}

\doi{doi: 10.4208/cicc.2026.xxx.xx}
\publishedyear{2026}
\volume{1}
\issue{1}
\receiveddate{xxx}
\revisiondate{xxx}
\accepteddate{xxx}
\onlinedate{xxx}
\publisheddate{xxx}

\title[Dissipaton-embedded quantum master equation for nonlinear system–bath coupling]
{Dissipaton-embedded quantum master equation  for nonlinear system-bath coupling}
\author[1]{Zi-Fan Zhu}
\author[1,*]{Yao Wang}
\author[1,2]{Rui-Xue Xu}
\author[1,2]{YiJing Yan}

\affil[1]{\textit{Hefei National Research Center for Physical Sciences at the Microscale, University of Science and Technology of China, Hefei 230026, China} }
\affil[2]{\textit{Department of Chemical
Physics, University of Science and Technology of China, Hefei 230026, China} 
\protect\vspace{1em}}

\affil[*]{Corresponding authors: wy2010@ustc.edu.cn}

\begin{document}

\twocolumn[{
  \vspace*{0.5em}
  \maketitle
  \thispagestyle{firstpage}
  \label{firstpage}

\begin{abstract}
The dissipaton theory was proposed more than a decade ago as a statistical quasiparticle theory for open quantum systems and was soon extended to nonlinear system–bath coupling. Although numerical results provided strong evidence for its validity, a rigorous mathematical foundation, particularly for nonlinear system–bath coupling, was established only recently. On this basis, we formulate the dissipaton-embedded quantum master equation for nonlinear system–bath coupling. This in-principle exact master equation describes the joint dynamics of the reduced system and all dissipaton degrees of freedom.

\end{abstract}

}] 

\textbf{Introduction.} An open quantum system consists of the degrees of freedom of interest coupled to an environment, or bath. Tracing over the bath gives the reduced density operator of the system \cite{Wei21,Bre02}. When bath correlations persist on the time scale of the system, the reduced dynamics can be non-Markovian, retaining a memory of earlier system–bath interactions \cite{Bre16021002,Dev17015001,Xu26021002}.
In 2014, Yan introduced \emph{dissipatons} as statistical quasiparticles that resolve bath fluctuations into modes, each associated with a term in the bath correlation function \cite{Yan14054105}. The dissipaton equation of motion (DEOM) organizes the dynamics into a hierarchy of density operators. 
Its lowest tier gives the reduced system operator, and higher tiers encode correlations with the bath. 
For linear system–bath coupling to a Gaussian bath, DEOM is identical to the famous hierarchical equations of motion (HEOM) \cite{Tan89101,Tan906676,Yan04216,Xu05041103,Tan20020901}. 

The DEOM was soon extended from linear to quadratic bath coupling \cite{Xu17395,Xu18114103}, and numerical benchmarks supported this extension \cite{Che23074102}. Recently, a rigorous mathematical foundation for nonlinear (quadratic and higher-order) system–bath coupling in a Gaussian environment has only recently been established \cite{Su26Arxiv2609_24603}.  
By constructing dissipatons in the thermofield space of the bath and applying a recursive Wick theorem, Ref.~\cite{Su26Arxiv2609_24603} derived the hierarchy for polynomial couplings directly from the Liouville equation.
This approach also provides an explicit treatment of exponential system--bath coupling~\cite{Cha25Arxiv2510_25133}.

For linear system–bath coupling, researchers recast the DEOM hierarchy as a single dissipaton-embedded quantum master equation (DQME) \cite{Li23214110}. The original formulation treats bosonic dissipatons as generalized Brownian particles. A subsequent formulation in second quantization describes transitions between dissipaton configurations through creation and annihilation operators for either bosonic or fermionic baths \cite{Li24032620, Su254107}. 
The present work extends the bosonic DQME to nonlinear system–bath coupling. The resulting exact master equation governs the joint evolution of the reduced system and all dissipaton degrees of freedom, retaining the system–bath correlations encoded in the DEOM hierarchy.

\textbf{System--bath model and dissipaton decomposition.} The construction of DQME begins with the collective bath coordinate that enters
the system--bath interaction. Throughout, we set $\hbar=1$ and write
$\beta=1/(k_{\rm B}T)$, where $k_{\rm B}$ is the Boltzmann constant and
$T$ is the bath temperature. The total Hamiltonian reads
\begin{equation}
    H_{\rm T}=H_{\rm S}+H_{\rm B}+H_{\rm SB},
\end{equation}
where $H_{\rm S}$ and $H_{\rm B}$ are the Hamiltonians of the isolated
system and the bare bosonic bath, respectively, and $H_{\rm SB}$
describes their interaction. 

For a single coupling channel, the
interaction takes the form 
\begin{equation}
H_{\rm SB}=\hat Q\,g(\hat F), 
\end{equation}
where $\hat Q$
is a system operator and $\hat F$ is a Hermitian collective bath
coordinate \cite{Zhu26HigherOrderBath}. The function $g$ specifies how this coordinate couples to
the system. Linear coupling corresponds to $g(\hat F)=\alpha_1 \hat F$, with $\alpha_1$ being the
coupling strength. 
Quadratic and higher-order couplings involve powers of the same bath
coordinate. The interaction generalizes directly to multiple coupling
channels as $H_{\rm SB}=\sum_a \hat Q_a\,g_a(\hat F_a)$, where $a$
labels the channels. For simplicity, the DQME is derived below for a
single channel.

For a Gaussian bath, the equilibrium fluctuations of $\hat F$ are fully
specified by its mean and its two-time correlation function. Taking
$\langle\hat F\rangle_{\rm B}=0$, the bath input to the dynamics is
therefore
\begin{equation}
    C(t)=\langle\hat F(t)\hat F(0)\rangle_{\rm B},
    \qquad
    \hat F(t)=e^{iH_{\rm B}t}\hat F e^{-iH_{\rm B}t}.
\end{equation}
Here the time evolution is generated by the bare bath Hamiltonian, and
the expectation value of an arbitrary bath operator $\hat O$ is taken
in the thermal equilibrium state,
\begin{equation}
    \langle\hat O\rangle_{\rm B}
    \equiv
    \frac{{\rm Tr}_{\rm B}\!\left(\hat O e^{-\beta H_{\rm B}}\right)}
         {{\rm Tr}_{\rm B}\!\left(e^{-\beta H_{\rm B}}\right)}.
\end{equation}
The time dependence of $C(t)$ encodes the bath memory. For nonlinear
coupling, the dynamics also involves higher-order correlations of
$\hat F$. Wick's theorem expresses these correlations in terms of
pair correlations, so that preserving $C(t)$ also preserves the bath
statistics needed to treat $g(\hat F)$ \cite{Su26Arxiv2609_24603}.

The dissipaton decomposition  represents $\hat F$ as a sum of dissipaton coordinate
operators $\{\hat f_k\}$~\cite{Yan14054105,Li24032620},
\begin{equation}
    \label{map1}
    \hat F \longmapsto \sum_{k=1}^{K}\hat f_k .
\end{equation}
Each dissipaton carries one component $c_k(t)$ of the bath correlation
function. The components are chosen so that dissipatons with distinct
indices have vanishing pair correlations,
\begin{equation}
    \langle\hat f_k(t)\hat f_{k'}(0)\rangle_{\rm D}
    =\delta_{kk'}c_k(t),
\end{equation}
where $\langle\cdots\rangle_{\rm D}$ denotes the expectation value in
the dissipaton vacuum state~\cite{Su26Arxiv2609_24603}. The condition that the
decomposition preserve the original bath correlation function then reads
\begin{equation}
    \label{condition}
    C(t)
    =\sum_{k,k'=1}^{K}
      \langle\hat f_k(t)\hat f_{k'}(0)\rangle_{\rm D}
    =\sum_{k=1}^{K}c_k(t).
\end{equation}
In this representation, the temperature dependence of the physical bath
is carried by the functions $c_k(t)$.

For $t\geq 0$, each component is represented by a single exponential,
\begin{equation}
    c_k(t)=\eta_k e^{-\gamma_k t},
\end{equation}
where $\eta_k$ is its amplitude  and $\gamma_k$ its exponent. They are both generally complex-valued. For a
given bath spectral density and temperature, the amplitudes and
exponents can be obtained by applying e.g. the Prony decomposition to
$C(t)$~\cite{Che22221102, Zha25214111b}. In numerical calculations, the number of
terms $K$ is chosen to reproduce $C(t)$ to the required accuracy.

The exponents are taken to be real or to occur in complex-conjugate
pairs~\cite{Yan14054105}. For each $k$, the partner index $\bar{k}$ is
defined by 
\begin{equation}
\gamma_{\bar{k}}=\gamma_k^{*}.
\end{equation}
Writing
$\gamma_k=\kappa_k+i\omega_k$ gives $
    c_k(t)=\eta_k e^{-\kappa_k t}e^{-i\omega_k t}$.
Thus $\kappa_k$ gives the decay rate, while $\omega_k$ gives the 
oscillation frequency. For a real exponent, $\bar{k}=k$ and the mode decays without
oscillation. For a complex-conjugate pair, the two modes have the same
decay rate and opposite oscillation frequencies.

\textbf{DEOM for nonlinear system--bath coupling.}
Consider the coupling $H_{\rm SB}=\hat Q\,g(\hat F)$ introduced above,
with 
\begin{equation}
g(u)=\sum_{r=0}^{R}\alpha_r u^r
\end{equation}
of arbitrary finite degree $R$.
The constant term can equivalently be included in $H_{\rm S}$. For
$t\geq0$, the forward and reverse bath correlations have amplitudes
$\eta_k^{>}=\eta_k$ and $\eta_k^{<}=\eta_{\bar k}^{*}$, respectively.
Indeed, relabeling the exponents in $C(-t)=[C(t)]^*$ gives
$C(-t)=\sum_k\eta_k^{<}e^{-\gamma_k t}$. At equal times,
\begin{equation}
 \sum_k\eta_k^{>}=\sum_k\eta_k^{<}
 =C(0)=\langle\hat F^2\rangle_{\rm B}.
 \label{eq:equal-time}
\end{equation}

Following the higher-order DEOM construction~\cite{Zhu26HigherOrderBath},
let $\mathbf n=(n_1,\ldots,n_K)$ denote the dissipaton occupations,
with $n_k=0,1,\ldots$ and tier $n=\sum_k n_k$. The dissipaton density
operators are
\begin{equation}
 \rho_{\mathbf n}(t)={\rm Tr}_{\rm B}\!\left[
 \left(\prod_k\hat f_k^{n_k}\right)^{\!\circ}\rho_{\rm T}(t)\right].
 \label{eq:dissipaton-moments}
\end{equation}
The circle denotes the irreducible product with respect to the bare
Gaussian bath~\cite{Yan14054105}, meaning that the dissipaton operators
are normal ordered with respect to the thermofield vacuum~\cite{Su26Arxiv2609_24603}. It is symmetric in its bosonic factors. The zeroth
tier is $\rho_{\mathbf 0}=\rho_{\rm S}$, and all higher tiers vanish
initially for a factorized state with the bath in thermal equilibrium.

To insert a power of $\hat F$, replace $\rho_{\rm T}$ in
Eq.~(\ref{eq:dissipaton-moments}) by any composite operator $A$ and
denote its hierarchy by $a=\{a_{\mathbf n}\}$. The generalized Wick
theorem gives the hierarchies of $\hat F A$ and $A\hat F$ through
the two maps~\cite{Zhu26HigherOrderBath}
\begin{equation}
 (\mathcal F_{\sigma}a)_{\mathbf n}
 =\sum_k\!\left(a_{\mathbf n+\mathbf e_k}
       +n_k\eta_k^{\sigma}a_{\mathbf n-\mathbf e_k}\right),
 \quad \sigma\in\{>,<\}.
 \label{eq:force-insertions}
\end{equation}
Here $\mathbf e_k$ is the unit vector for mode $k$, and terms with a negative
occupation vanish. The first term adds an uncontracted dissipaton.
The second contracts it with one of the $n_k$ existing factors.
Since the rule holds for any $A$, the hierarchies of
$\hat F^r\rho_{\rm T}$ and $\rho_{\rm T}\hat F^r$ are
$\mathcal F_{>}^{r}\rho$ and $\mathcal F_{<}^{r}\rho$, respectively,
where $\rho=\{\rho_{\mathbf n}\}$. The Wick rule is applied recursively, so each new factor of $\hat F$
can contract with factors introduced at earlier steps.

Projecting $\dot\rho_{\rm T}=-i[H_{\rm T},\rho_{\rm T}]$ onto
Eq.~(\ref{eq:dissipaton-moments}) gives the arbitrary-order DEOM ~\cite{Zhu26HigherOrderBath},
\begin{equation}
\begin{array}{@{}l@{}}
 \displaystyle
 \dot\rho_{\mathbf n}=-i[H_{\rm S},\rho_{\mathbf n}]
       -\Big(\sum_k n_k\gamma_k\Big)\rho_{\mathbf n}\\[2pt]
 \displaystyle\qquad
 -i\sum_{r=0}^{R}\alpha_r
  \left[\hat Q(\mathcal F_{>}^{r}\rho)_{\mathbf n}
            -(\mathcal F_{<}^{r}\rho)_{\mathbf n}\hat Q\right].
\end{array}
\label{eq:nonlinear-deom}
\end{equation}
The second term in the first line follows from free bath evolution.
A monomial of degree $r$ connects tiers whose differences belong
to $\{-r,-r+2,\ldots,r\}$. Its degree therefore fixes the range of
direct tier couplings.

\textbf{DQME for nonlinear system--bath coupling.}
The hierarchy in Eq.~(\ref{eq:nonlinear-deom}) admits two equivalent
representations. In dissipaton coordinates its tier shifts become
differential operators. In the occupation basis the same shifts are
generated by bosonic creation and annihilation operators.

\emph{Coordinate representation.}
Associate a real dimensionless coordinate $x_k$ with each dissipaton
component~\cite{Li23214110}, and set
\begin{equation}
 \zeta_k^2=\frac{\eta_k+\eta_{\bar k}^{*}}{2},\qquad
 \xi_k=\frac{\eta_k-\eta_{\bar k}^{*}}{2i\zeta_k}.
 \label{eq:hermite-scales}
\end{equation}
A square-root branch is fixed for each nonzero $\zeta_k$.
The Hermite polynomials $H_m$ are defined by
\begin{equation}
 H_m(y)=(-1)^m e^{y^2}
 \frac{{\rm d}^{m}}{{\rm d}y^{m}}e^{-y^2},
 \qquad m=0,1,\ldots .
 \label{eq:physicists-hermite}
\end{equation}
With the reference Gaussian $w(x)=e^{-x^2/2}/\sqrt{2\pi}$, choose
the one-mode basis functions
\begin{equation}
 \varphi_m(x)=\frac{w(x)}{2^{m/2}m!}
                  H_m\!\left(\frac{x}{\sqrt{2}}\right).
 \label{eq:hermite-basis}
\end{equation}
The argument $x/\sqrt{2}$ converts the Hermite weight $e^{-y^2}$
to $e^{-x^2/2}$. The corresponding orthogonality relation is
\begin{equation}
 \int_{-\infty}^{\infty}{\rm d}x\,
 2^{-l/2}H_l(x/\sqrt{2})\,\varphi_m(x)=\delta_{lm}.
 \label{eq:hermite-duality}
\end{equation}

Let $a=\{a_{\mathbf n}\}$ be any hierarchy of system operators.
The symbol $\mathcal T$ denotes the linear map from this entire
sequence to a single operator-valued function of
$\mathbf x=(x_1,\ldots,x_K)$,
\begin{equation}
 (\mathcal T a)(\mathbf x)
 \equiv\sum_{\mathbf n}a_{\mathbf n}
       \prod_{k=1}^{K}\frac{\varphi_{n_k}(x_k)}{\zeta_k^{n_k}} .
 \label{eq:hierarchy-transform}
\end{equation}
In particular, $\hat\rho(\mathbf x,t)=(\mathcal T\rho)(\mathbf x,t)$
is the image of the dissipaton density operators
$\rho=\{\rho_{\mathbf n}(t)\}$. Orthogonality recovers every member,
\begin{equation}
 \rho_{\mathbf n}(t)=\int {\rm d}\mathbf x\,\hat\rho(\mathbf x,t)
       \prod_k\!\left(\frac{\zeta_k}{\sqrt{2}}\right)^{n_k}
                   H_{n_k}\!\left(\frac{x_k}{\sqrt{2}}\right),
 \label{eq:hermite-pair}
\end{equation}
where ${\rm d}\mathbf x=\prod_k{\rm d}x_k$. At $\mathbf n=\mathbf 0$ this gives
$\rho_{\rm S}(t)=\int{\rm d}\mathbf x\,\hat\rho(\mathbf x,t)$.
The expansion and its inverse are understood on hierarchies for
which the indicated moments exist.

Write $\partial_k$ for differentiation with respect to $x_k$ at
fixed time and fixed remaining coordinates. The Hermite recurrences give
\begin{equation}
 \left(x+\frac{{\rm d}}{{\rm d}x}\right)\varphi_m=\varphi_{m-1},
 \qquad
 -\frac{{\rm d}\varphi_m}{{\rm d}x}=(m+1)\varphi_{m+1},
 \label{eq:hermite-recurrences}
\end{equation}
with $\varphi_{-1}=0$. Relabeling the sums in
Eq.~(\ref{eq:hierarchy-transform}) then gives the two parts of a Wick
insertion,
\begin{equation}
\begin{array}{@{}l@{}}
 \mathcal T\{a_{\mathbf n+\mathbf e_k}\}
   =\zeta_k(x_k+\partial_k)\mathcal T a,\\[2pt]
 \mathcal T\{n_k a_{\mathbf n-\mathbf e_k}\}
   =-\zeta_k^{-1}\partial_k\mathcal T a .
\end{array}
\label{eq:tier-shifts}
\end{equation}
Their combination in Eq.~(\ref{eq:force-insertions}) represents a
single insertion of $\hat f_k$ from the left or the right. Since
$\eta_k^{>}=\eta_k$ and $\eta_k^{<}=\eta_{\bar k}^{*}$, it acts as
\begin{equation}
 \zeta_k(x_k+\partial_k)-\frac{\eta_k^\sigma}{\zeta_k}\partial_k
 =\left\{\begin{array}{ll}
  \zeta_k x_k-i\xi_k\partial_k, & \sigma=~>,\\
  \zeta_k x_k+i\xi_k\partial_k, & \sigma=~<.
 \end{array}\right.
 \label{eq:single-coordinate-insertion}
\end{equation}
The same recurrences give
$\partial_k(x_k+\partial_k)\varphi_{n_k}=-n_k\varphi_{n_k}$,
so the free bath drift is represented by
\begin{equation}
\hat\Gamma=\sum_k\gamma_k\partial_k(x_k+\partial_k).
\end{equation}

To proceed, define
\begin{equation}
 X=\sum_k\zeta_k x_k,\qquad \hat D=\sum_k\xi_k\partial_k .
 \label{eq:force-coordinate}
\end{equation}
Equation~(\ref{eq:single-coordinate-insertion}) intertwines the
insertion maps with $X-i\hat D$ and $X+i\hat D$ for every hierarchy
$a$. Applying this identity successively retains the contractions
between different factors of $\hat F$. For any $r\geq 0$,
\begin{equation}
\begin{array}{@{}l@{}}
 \mathcal T(\mathcal F_{>}^{r}\rho)=(X-i\hat D)^r\hat\rho,\\[2pt]
 \mathcal T(\mathcal F_{<}^{r}\rho)=(X+i\hat D)^r\hat\rho .
\end{array}
\label{eq:repeated-insertions}
\end{equation}
The equal-time identity in Eq.~(\ref{eq:equal-time}) implies
\begin{equation}
 [\hat D,X]=\sum_k\zeta_k\xi_k
 =\frac{1}{2i}\sum_k(\eta_k-\eta_{\bar k}^{*})=0 .
 \label{eq:collective-commutation}
\end{equation}
Thus for the polynomial $g(u)=\sum_{r=0}^{R}\alpha_r u^r$,
\begin{equation}
 g(X\mp i\hat D)=\sum_{m=0}^{R}(\mp i)^m h_m(X)\hat D^m,
 \label{eq:operator-polynomial}
\end{equation}
where $h_m(X)=(1/m!)\,{\rm d}^m g(X)/{\rm d}X^m$.
With $\hat X_{\pm}\equiv X\pm i\hat D$, substitution into Eq.~(\ref{eq:nonlinear-deom}) gives 
the nonlinear DQME,
\begin{equation}\label{eq:nonlinear-dqme}
\begin{split}
 \frac{{\rm d}\hat\rho}{{\rm d}t}
 =-i[H_{\rm S},\hat\rho]+\hat\Gamma\hat\rho
 -i\hat Q\,[g(\hat X_{-})\hat\rho]
 +i\big[g(\hat X_{+})\hat\rho\big]\hat Q .
 \end{split}
\end{equation}
The time derivative is taken at fixed $\mathbf x$. Even powers of
$\hat D$ multiply commutators with $\hat Q$, and odd powers multiply anticommutators.

\emph{Occupation representation.}
Introduce one auxiliary bosonic Fock space $\mathcal H_{{\rm D},k}$
for each component of the bath correlation function~\cite{Li24032620}.
Its orthonormal number states $|n_k\rangle$, $n_k=0,1,\ldots$, count
the occupation of that dissipaton. The joint occupation basis and the
space of the embedded density tensor are
\begin{equation}
\begin{array}{@{}l@{}}
 \displaystyle
 \mathcal H_{\rm D}=\bigotimes_{k=1}^{K}\mathcal H_{{\rm D},k},
 \qquad |\mathbf n\rangle=\bigotimes_{k=1}^{K}|n_k\rangle,\\[3pt]
 \displaystyle
 \widetilde\rho(t)\in
 (\mathcal H_{\rm S}\otimes\mathcal H_{\rm S}^{*})
       \otimes\mathcal H_{\rm D}.
\end{array}
\label{eq:sq-space}
\end{equation}
Here $\mathcal H_{\rm S}\otimes\mathcal H_{\rm S}^{*}$ is the system
Liouville space. Each coefficient of $|\mathbf n\rangle$ is a system
operator with a bra and a ket index. Each dissipaton contributes one
occupation index, inherited from $\rho_{\mathbf n}$.

To pass from coordinates to occupations, identify the one-mode basis
function $\varphi_m$ with $|m\rangle/\sqrt{m!}$ and extend the map
linearly to products over $k$. Denote this change of representation by
$\mathcal U$. Equation~(\ref{eq:hierarchy-transform}) then gives
\begin{equation}
 \widetilde\rho(t)=\mathcal U\hat\rho
 =\sum_{\mathbf n}
 \frac{\rho_{\mathbf n}(t)\otimes|\mathbf n\rangle}
      {\prod_k\zeta_k^{n_k}\sqrt{n_k!}} .
 \label{eq:sq-tensor}
\end{equation}
Projection onto the auxiliary vacuum gives
$(\mathrm{id}_{\rm S}\otimes\langle\mathbf 0|)\widetilde\rho
=\rho_{\rm S}$, where $\mathrm{id}_{\rm S}$ leaves system operators
unchanged. For a factorized thermal preparation,
$\widetilde\rho(0)=\rho_{\rm S}(0)\otimes|\mathbf 0\rangle$.

The factorial in Eq.~(\ref{eq:sq-tensor}) fixes the ladder
normalization. Under $\mathcal U$, the two recurrences in
Eq.~(\ref{eq:hermite-recurrences}) become
\begin{equation}
\begin{array}{@{}l@{}}
 (x_k+\partial_k)\ \longmapsto\ \hat b_k,\qquad
 \hat b_k|\mathbf n\rangle=\sqrt{n_k}\,|\mathbf n-\mathbf e_k\rangle,\\[3pt]
 -\partial_k\ \longmapsto\ \hat b_k^\dagger,\qquad\qquad\,
 \hat b_k^\dagger|\mathbf n\rangle
       =\sqrt{n_k+1}\,|\mathbf n+\mathbf e_k\rangle .
\end{array}
\label{eq:sq-ladders}
\end{equation}
Thus $[\hat b_k,\hat b_l^\dagger]=\delta_{kl}$, and
$x_k$ is represented by $\hat b_k+\hat b_k^\dagger$. The free bath
generator becomes $-\sum_k\gamma_k\hat N_k$, with
$\hat N_k=\hat b_k^\dagger\hat b_k$.

Let $\hat B_{>}$ and $\hat B_{<}$ denote the occupation-space
representatives of $X-i\hat D$ and $X+i\hat D$, respectively.
Transforming Eq.~(\ref{eq:single-coordinate-insertion}) gives both,
\begin{equation}
 \hat B_\sigma=\sum_k\left(
    \zeta_k\hat b_k+\frac{\eta_k^\sigma}{\zeta_k}\hat b_k^\dagger
                   \right),\qquad \sigma\in\{>,<\}.
 \label{eq:sq-insertions}
\end{equation}
In particular, the creation amplitudes are
$\eta_k^{>}/\zeta_k=\zeta_k+i\xi_k$ and
$\eta_k^{<}/\zeta_k=\zeta_k-i\xi_k$. For each power in the coupling,
Eq.~(\ref{eq:repeated-insertions}) therefore becomes
\begin{equation}
 \mathcal U\mathcal T(\mathcal F_\sigma^r\rho)
       =\hat B_\sigma^r\widetilde\rho .
 \label{eq:sq-powers}
\end{equation}
The second-quantized DQME follows by summing these powers with
their original coefficients $\alpha_r$,
\begin{equation}
\begin{array}{@{}l@{}}
 \displaystyle
 \frac{{\rm d}\widetilde\rho}{{\rm d}t}
 =-i[H_{\rm S},\widetilde\rho]
     -\sum_k\gamma_k\hat N_k\widetilde\rho
 -i\hat Q\,[g(\hat B_{>})\widetilde\rho]
 +i\big[g(\hat B_{<})\widetilde\rho\big]\hat Q .
\end{array}
\label{eq:nonlinear-sq-dqme}
\end{equation}
Both $g(\hat B_\sigma)$ act on the auxiliary occupation index.
The position of $\hat Q$ specifies left or right multiplication on
the system operator. For $g(u)=\alpha_1u$, the interaction reduces to
\begin{equation}
 -i\alpha_1\sum_k\zeta_k(\hat b_k+\hat b_k^\dagger)
                  [\hat Q,\widetilde\rho]
 +\alpha_1\sum_k\xi_k\hat b_k^\dagger
                  \{\hat Q,\widetilde\rho\},
\label{eq:sq-linear-limit}
\end{equation}
which recovers the linear bosonic DQME in second
quantization~\cite{Li24032620}.

The symmetric scaling in Eq.~(\ref{eq:hermite-scales}) assumes
$\zeta_k\ne0$. If a component has $\zeta_k=0$, choose a nonzero
scale $s_k$ for that component in the forward and inverse transforms
and in Eq.~(\ref{eq:sq-tensor}). Its left or right coordinate
insertion then reads
$s_k(x_k+\partial_k)-(\eta_k^\sigma/s_k)\partial_k$, and its
contribution to $\hat B_\sigma$ is
$s_k\hat b_k+(\eta_k^\sigma/s_k)\hat b_k^\dagger$.
The two representations are equivalent to the full hierarchy for
any finite polynomial degree. The operator-function notation also
extends to a power series whenever convergence permits termwise
transformation.

\textbf{Discussion.} Below are some remarks:

\emph{(i) Range of applicability.}
The single-channel construction extends directly in three directions.
\begin{itemize}
\item For $H_{\rm SB}=\sum_a\hat Q_a\,g_a(\hat F_a)$, each channel
contributes its own left and right insertions. For correlated bath
coordinates these insertions are fixed by the full correlation matrix
$C_{ab}(t)=\langle\hat F_a(t)\hat F_b(0)\rangle_{\rm B}$.
\end{itemize}
\begin{itemize}
\item Both $H_{\rm S}$ and $\hat Q_a$ may depend explicitly on time.
An applied field $E(t)$ acting on a dipole
$\hat\mu_0+\hat\mu_1\hat F$, for example, adds $-E(t)\hat\mu_0$
to $H_{\rm S}(t)$ and $-E(t)\hat\mu_1$ to the coefficient of
$\hat F$. A bath dipole proportional to $\hat F$ gives direct bath
driving, as in Fano interference~\cite{Zha15024112}. A transition
dipole that varies with $\hat F$ gives Herzberg--Teller non-Condon
coupling~\cite{Zha16204109}.
\end{itemize}
\begin{itemize}
\item A purely bath term $c_r(t)\hat F^r$ is included by setting
$\hat Q=c_r(t)\hat I_{\rm S}$ and $g(u)=u^r$, where
$\hat I_{\rm S}$ is the system identity. Anharmonic terms of degree
$r\geq3$ can thus be propagated explicitly while $H_{\rm B}$
defines the Gaussian reference bath.
\end{itemize}

\emph{(ii) Conjugation symmetry.}
Let $\bar{\mathbf n}$ denote the occupation vector with
$\bar n_k=n_{\bar k}$. Since $\gamma_{\bar k}=\gamma_k^*$ and
$(\eta_k^{>})^*=\eta_{\bar k}^{<}$, Hermitian conjugation followed
by $k\leftrightarrow\bar k$ interchanges the left and right insertions
in Eq.~(\ref{eq:force-insertions}). For a Hermitian Hamiltonian,
Eq.~(\ref{eq:nonlinear-deom}) therefore preserves
\begin{equation}
 \rho_{\mathbf n}(t)^\dagger=\rho_{\bar{\mathbf n}}(t)
 \label{eq:paired-ddos}
\end{equation}
whenever it holds initially, as it does for a factorized thermal
preparation. Each self-paired configuration, $\mathbf n=\bar{\mathbf n}$,
has a Hermitian density operator. For distinct paired configurations,
one member suffices to reconstruct the other. In a coordinate
convention with $\zeta_{\bar k}=\zeta_k^*$, and $\xi_{\bar k}=\xi_k^*$, the same symmetry reads
$\hat\rho(\mathbf x,t)^\dagger=\hat\rho(\bar{\mathbf x},t)$, with
$\bar x_k=x_{\bar k}$.

\emph{(iii) Exact factorization.}
The coordinate distribution admits a marginal--conditional
factorization analogous to that used for wavefunctions~\cite{Abedi10ExactFactorization}.
Define $P\equiv{\rm Tr}_{\rm S}\hat\rho$ and, wherever $P\ne0$,
$\hat W\equiv\hat\rho/P$. Then
\begin{equation}
 \hat\rho=P\hat W,\qquad {\rm Tr}_{\rm S}\hat W=1,\qquad
 \rho_{\rm S}(t)=\!\!\int\!\! {\rm d}\mathbf x\,P\hat W .
\label{eq:factorization}
\end{equation}
For the single channel, put $q={\rm Tr}_{\rm S}(\hat Q\hat W)$ and
$\hat X_\pm=X\pm i\hat D$. Taking the system trace of
Eq.~(\ref{eq:nonlinear-dqme}) gives
\begin{equation}
 \frac{{\rm d}P}{{\rm d}t}
 =\hat\Gamma P-i\big[g(\hat X_-)-g(\hat X_+)\big](Pq).
 \label{eq:marginal}
\end{equation}
Write $\mathcal K$ for the full DQME generator,
\({\rm d}\hat\rho/{\rm d}t=\mathcal K\hat\rho\). The product rule
then gives the conditional evolution at fixed $\mathbf x$,
\begin{equation}
 \frac{{\rm d}\hat W}{{\rm d}t}
 =\frac{1}{P}\left[\mathcal K(P\hat W)
                  -\hat W\frac{{\rm d}P}{{\rm d}t}\right].
 \label{eq:conditional}
\end{equation}
The conditional expectation $q$ drives the marginal, while derivatives
of $P$ in $\mathcal K(P\hat W)$ enter the conditional evolution.
Their product reconstructs the full DQME. Near zeros of $P$,
$\hat\rho$ itself remains the regular variable for propagation.

The two factors may require different spatial resolutions. If $P$
is localized and $\hat W$ varies slowly across the region that
contributes to $\rho_{\rm S}$, a fine grid could resolve the scalar
marginal while a coarser grid carries the conditional system
operators. Interpolating $\hat W$ between these nodes would reduce
the number of full system matrices stored and propagated.
The attainable saving can be tested by comparing cost at fixed
accuracy in $\rho_{\rm S}(t)$ and the relevant bath moments against
direct propagation of $\hat\rho$.

An important direction for future development is to exploit the underlying physical structure of the dissipaton degrees of freedom to achieve more efficient representations. Tensor-network methods, particularly matrix product state (MPS) and related renormalization-group approaches, may provide physically informed compression beyond conventional truncation~\cite{Ye25120901,gu2026nonadiabaticrenormalizationgroupstrongly}. Complementary approaches based on the Davydov ansatz and neural quantum states, as well as digital quantum algorithms, offer further possibilities for extending DQME to more complex systems~\cite{Shi25224103,CaoL140301,Cao26194101,Li24032620}. Moreover, recent advances in memory kernel coupling theory \cite{Liu25148001}, including projection-based stabilization and continued-fraction representations, suggest complementary strategies for reducing computational costs while preserving long-time stability~\cite{Liu26StableMemoryKernel}. These developments may ultimately lead to more efficient and physically transparent approaches to non-Markovian quantum dynamics.

\begin{acknowledgments}
Support from the National Natural Science Foundation of China (22373091 and 22573099) is gratefully acknowledged. 
\end{acknowledgments}

\begin{Notes}
The authors declare no competing financial interest.
\end{Notes}

% \bibliographystyle{cicc}
% \bibliography{bibrefs}

\label{lastpage}
\end{document}